\documentclass[10pt, aps, prd, amsmath, amssymb, floats, floatfix, twocolumn, notitlepage, superscriptaddress, nofootinbib, showpacs, longbibliography]{revtex4-2}
\usepackage{orcidlink}
\usepackage[T1]{fontenc}
\usepackage[utf8]{inputenc}
\usepackage{lmodern}
\usepackage{microtype}
\usepackage{amsmath,amssymb,amsfonts,bm,mathtools}
\usepackage{mathrsfs}
\usepackage{booktabs}
\usepackage{xcolor}
\usepackage{orcidlink}
\usepackage{hyperref}
\hypersetup{
  colorlinks=true,
  linkcolor=blue!45!black,
  citecolor=blue!45!black,
  urlcolor=blue!45!black,
  pdftitle={Non-minimal unimodular inflation revisited: Swampland diagnostics and CMB constraints}
}
\hypersetup{colorlinks=true,linkcolor=black,citecolor=black,urlcolor=blue}
\renewcommand{\arraystretch}{1.18}
\allowdisplaybreaks
\providecommand{\diff}{\mathrm{d}}
\newcommand{\mpl}{M_{\rm P}}
\newcommand{\epsV}{\epsilon_V}
\newcommand{\etaV}{\eta_V}

\newcommand{\As}{A_s}
\newcommand{\ns}{n_s}
\newcommand{\rr}{r}
\newcommand{\alphas}{\alpha_s}
\newcommand{\SWGC}{\mathrm{SWGC}}
\newcommand{\SSWGC}{\mathrm{SSWGC}}
\newcommand{\FRDSSC}{\mathrm{FRDSSC}}

\begin{document}

\title{Non-minimal unimodular inflation revisited: Swampland diagnostics and CMB constraints}

\author{Saeed Noori Gashti\orcidlink{0000-0001-7844-2640}}
\email[Email: ]{sn.gashti@du.ac.ir}
\affiliation{School of Physics, Damghan University, P. O. Box 3671641167, Damghan, Iran}
\affiliation{Center for Theoretical Physics, Khazar University, 41 Mehseti Street, Baku, AZ1096, Azerbaijana}
\author{Mohammad Reza Alipour\orcidlink{0000-0001-8074-7865}}
\email[Email: ]{mr.alipour@stu.umz.ac.ir}
\affiliation{School of Physics, Damghan University, P. O. Box 3671641167, Damghan, Iran}
\affiliation{Center for Theoretical Physics, Khazar University, 41 Mehseti Street, Baku, AZ1096, Azerbaijana}
\author{Mohammad Ali S. Afshar\orcidlink{0009-0001-3133-5992}}
\email[Email: ]{m.a.s.afshar@gmail.com}
\affiliation{Department of Theoretical Physics, Faculty of Basic Sciences, University of Mazandaran, P. O. Box 47416-95447, Babolsar, Iran}
\affiliation{School of Physics, Damghan University, P. O. Box 3671641167, Damghan, Iran}
\affiliation{Center for Theoretical Physics, Khazar University, 41 Mehseti Street, Baku, AZ1096, Azerbaijana}
\author{Behnam Pourhassan\orcidlink{0000-0003-1338-7083}}
\email[Email: ]{b.pourhassan@du.ac.ir}
\affiliation{School of Physics, Damghan University, P. O. Box 3671641167, Damghan, Iran}
\affiliation{Center for Theoretical Physics, Khazar University, 41 Mehseti Street, Baku, AZ1096, Azerbaijana}
\author{\.{I}zzet Sakall{\i}\orcidlink{0000-0001-7827-9476}}
\email[Email: ]{izzet.sakalli@emu.edu.tr}
\affiliation{Physics Department, Eastern Mediterranean University, Famagusta 99628, North Cyprus via Mersin 10, T\"urkiye}
\author{Phongpichit Channuie\orcidlink{0000-0003-4489-1207}}
\email[Email: ]{channuie@gmail.com}
\affiliation{School of Science \& College of Graduate Studiea, Walailak University, Nakhon Si Thammarat, 80160, Thailand}

\begin{abstract}
We study non-minimal unimodular inflation in the Einstein frame; the effective potential takes the form $U(\varphi)=U_0(1+\xi\varphi^2/\mpl^2)^{-2}$. We consider the canonical field map and the inflationary observables directly from the Einstein-frame action. The exact potential slow-roll parameters, the leading scalar running, the scalar-amplitude normalization, the e-fold relation, and the end-of-inflation condition are obtained analytically. The model has a nontrivial structural tension: an $\epsilon_V=1$ exit requires $\xi>1/2$, whereas the near-origin slow-roll regime requires $\xi\ll1$ and gives $n_s\simeq1-8\xi$. We next test three distinct swampland diagnostics. The further-refined de Sitter conjecture (FRDSSC) introduces the independent conjectural parameters $a$, $b$, and $q$, while the scalar weak gravity conjecture (SWGC) and its strong form (SSWGC) contain no analogous tunable conjecture parameter in the single-field realization adopted here. All three criteria are evaluated using the canonically normalized Einstein-frame scalar. Analytic limits show that the SWGC is violated at the stationary point for every nonzero $\xi$, whereas the SSWGC at that point requires $\xi\gtrsim0.0643$. In the large-field limit both scalar weak-gravity diagnostics reduce to the same threshold $\xi\ge0.1$. Numerical scans then show that the observationally relevant ACT/P-ACT region does not satisfy either scalar weak-gravity criterion, while the FRDSSC can be satisfied for $a$ close to unity. Consequently, the minimal model can reproduce acceptable $(n_s,r)$ at isolated field values, but this does not constitute a complete 50-60 e-fold inflationary history or simultaneous satisfaction of all three conjectures.\\\\
\end{abstract}
\maketitle
\section{Introduction}\label{s1}
Recent joint analyses of the ACT-DR6, Planck 2018, and DESI BAO datasets \cite{2,3,4} have led to a substantial refinement of the parameter space allowed for inflationary models. The combined data provide stringent constraints on the primordial power spectrum, yielding
$
n_s = 0.9743 \pm 0.0034, \hspace{2pt} r < 0.038,
$
which correspond to an approximately $2\sigma$ upward shift in the preferred value of $n_s$ compared with the constraints obtained from Planck data alone \cite{2,3,4}. This shift has important implications for inflationary model
building: scenarios predicting comparatively low values of
$n_s$ are increasingly disfavored, whereas several extended
inflationary constructions remain compatible with the
ACT-era constraints \cite{6,7,8,9,10}..

The updated observational constraints also have direct implications for the post-inflationary evolution of the Universe. During reheating, the effective equation-of-state parameter $\omega_{\rm eff}$ plays a particularly important role in determining how the energy density evolves and, consequently, how primordial perturbations are transferred to the subsequent cosmological epochs. For $\alpha$-attractor models with hyperbolic tangent potentials, the combined ACT-DR6+Planck18+BAO analysis indicates that consistency with the data requires
$
\omega_{\rm eff} \geq 0.44,
$
a condition that does not arise when Planck 2018 data are considered in isolation \cite{15}. The relevance of this constraint extends beyond the reheating epoch itself. The evolution of primordial gravitational waves \cite{16,17}, as well as the generation and evolution of primordial magnetic fields through Faraday induction \cite{21,22}, depends explicitly on the effective post-inflationary equation of state. Consequently, the observational constraints on $\omega_{\rm eff}$
establish a direct connection between CMB measurements, tensor
perturbations, and primordial magnetogenesis \cite{23}.
These phenomenological developments can also be viewed from the perspective of the Swampland program, which is motivated by the expectation that quantum gravity imposes nontrivial consistency conditions on low-energy effective field theories. In this framework, the landscape denotes theories that can consistently arise from quantum gravity, whereas the swampland comprises effective theories that cannot be completed into a consistent quantum-gravitational description \cite{k,o,bbb}. Among the conjectures proposed within this program, the Weak Gravity Conjecture (WGC), dS conjecture, further-refined de Sitter conjecture, and the Trans-Planckian Censorship Conjecture (TCC) have attracted particular attention in discussions of inflation and dark-energy models \cite{50,51,52,53,54,55,56,57,58,59,60,61,62}. They provide complementary criteria for testing whether an apparently consistent low-energy cosmological scenario can remain compatible with a quantum theory of gravity.

An additional ingredient of recent inflationary constructions is the inclusion of quantum corrections to the inflaton potential. When these corrections are encoded through an anomalous scaling parameter $\gamma$, they can shift the predicted scalar spectral index toward the larger values preferred by the latest observations while maintaining a tensor-to-scalar ratio well below the present observational bound,
$
r \ll 0.038.
$
This behavior differs significantly from the predictions of minimal Starobinsky $R^2$ and Higgs-like inflationary models, which increasingly encounter tension with the upward shift in the preferred value of $n_s$ at approximately the $\gtrsim 2\sigma$ level.
Earlier approaches introduced a scale-invariant linear curvature coupling of the form $\phi^2\hat{R}$ in addition to the conventional quadratic curvature contribution \cite{18',19',20'}. Although related scale-invariant and Weyl-invariant
constructions can yield predictions close to those of
Starobinsky inflation, their detailed predictions depend
on the specific gravitational and scalar-field sector
\cite{18',19',20'}.

More recent constructions have therefore explored exponential rather than purely polynomial corrections to the curvature sector. The inclusion of exponential curvature terms can modify the inflationary dynamics without introducing the problematic inflaton-mass behavior associated with some polynomial extensions. At the same time, these corrections provide a controlled departure from the exact de Sitter limit characteristic of pure $\hat{R}^2$ gravity. Such a departure is phenomenologically important because the observed primordial spectrum is not exactly scale invariant. Recent work has explored several distinct extensions of
inflation in response to the ACT-era constraints. These
include exponential curvature extensions in Weyl gravity,
radiative corrections to Higgs-like inflation, polynomial
and higher-curvature deformations, string-motivated
derivative corrections, and modified-gravity models
\cite{21',22',23',24',25',26',27',28',29',30',31',
32',33',34',35',36',37'}. These developments therefore provide a useful framework for studying inflation together with its connections to reheating, primordial gravitational waves, and the generation of magnetic fields in the early Universe.

The present work is organized around three logically distinct levels of analysis: consistent model construction, phenomenological viability, and consistency with quantum-gravity-inspired conjectures. 
In Section~\ref{sec:intro}, we first clarify this distinction by separating genuine parameters of the inflationary model, such as $\xi$ and $U_0$, from the conjectural parameters $a$, $b$, and $q$ entering the FRDSSC, and from purely observational quantities such as the central values inferred from the ACT and Planck datasets. This separation is important because these three types of quantities play fundamentally different physical roles and should not be interpreted as interchangeable model parameters.

Section~\ref{sec:model} develops the exact Einstein-frame description of the minimal unimodular potential and derives the corresponding canonical field transformation. Establishing this field map at the outset removes possible ambiguities associated with the choice of field variable or conformal frame and provides the appropriate basis for all subsequent derivative-based consistency tests.
In Section~\ref{sec:observables}, we study the inflationary observables directly from the effective potential
$
U(\varphi)=U_0\left(1+\frac{\xi\varphi^2}{M_{\rm P}^2}\right)^{-2}.
$
The analysis includes the potential slow-roll parameters and the resulting predictions for the scalar spectral index $n_s$, the tensor-to-scalar ratio $r$, the leading running of the scalar spectral index $\alpha_s$, and the normalization of the scalar power spectrum. Expressing these quantities consistently in terms of the canonical field allows the observational predictions to be compared with the data without introducing an additional field-redefinition ambiguity.

Section~\ref{sec:efolds} then studies the duration of inflation and, in particular, the condition used to identify the end of the slow-roll regime. The analysis exposes an important tension between the small-$\xi$ regime required by the slow-roll treatment and the conventional end-of-inflation condition $\epsilon_V=1$. This point is essential because an acceptable value of the observables at a single pivot scale does not, by itself, guarantee the existence of a complete and self-consistent inflationary history lasting the required $50$-$60$ e-folds. 

The three Swampland diagnostics considered in this work, namely the FRDSSC, SWGC, and SSWGC, are formulated in canonical field variables in Section~\ref{sec:swampland}. Particular attention is paid to the different status of the parameters entering these criteria. In contrast to the FRDSSC, whose formulation involves the conjectural parameters $a$, $b$, and $q$, the SWGC and SSWGC do not introduce independently adjustable conjecture parameters. This distinction is retained throughout the analysis to avoid interpreting phenomenological freedom as evidence for quantum-gravity consistency.
A global numerical exploration of the model parameter space is presented in Section~\ref{sec:numerics}. Rather than testing the observational and Swampland requirements independently, we examine their simultaneous realization within the same parameter domain. The observational constraints from Planck 2018 and P-ACT-LB are imposed together with the three Swampland criteria, allowing us to determine whether a common region of parameter space survives all requirements simultaneously.

Finally, Sections~\ref{sec:discussion} and~\ref{sec:previous} collect the main results in a controlled comparative form and discuss their physical interpretation. In particular, we distinguish three issues that are sometimes conflated in inflationary model assessments: agreement with observations at an instantaneous pivot scale, the existence of a complete inflationary evolution over approximately $50$-$60$ e-folds, and simultaneous compatibility with the Swampland criteria. Keeping these requirements separate makes it possible to identify precisely where the minimal construction succeeds and where it encounters theoretical or dynamical limitations.

Section~\ref{sec:conclusion} summarizes the resulting no-go behavior of the minimal construction and discusses the directions in which the model would need to be extended in order to obtain a viable inflationary scenario satisfying the observational, dynamical, and quantum-gravity-inspired requirements simultaneously.
%%%%%%%%%%%%%%%%%%%%%%%%%%%%%%%
\section{Motivation and scope}
\label{sec:intro}
%%%%%%%%%%%%%%%%%%%%%%%%%%%%%%%
Inflationary model building involves several logically different tests. A candidate model must first admit a controlled accelerated phase, produce the observed scalar perturbation amplitude and spectral tilt, and respect the tensor bound. A separate question concerns whether the resulting low-energy effective theory is compatible with conjectural quantum-gravity diagnostics. The latter include several variants of the swampland program, among them the de Sitter family of conjectures and scalar generalizations of the weak gravity conjecture. These criteria should not be treated as experimental observables, and satisfying one of them does not establish ultraviolet completion.\\

The non-minimal unimodular construction considered here is attractive because its Jordan-frame scalar need not be supplied with an independent potential. The effective inflationary scale arises from the unimodular sector after the transformation to the Einstein frame. The original construction explicitly uses
\begin{equation}
f(\varphi)=\frac{\mpl^2}{2}\left(1+\frac{\xi\varphi^2}{\mpl^2}\right)
\label{eq:fjor}
\end{equation}
and obtains an Einstein-frame potential proportional to $(1+\xi\varphi^2/\mpl^2)^{-2}$, followed by a potential slow-roll analysis.\cite{Malekpour2024}
The aim of the present paper is deliberately narrower and more stringent. We ask whether the same model can simultaneously support a consistent inflationary history and satisfy the three diagnostics most often used in this context,
$
\FRDSSC,\quad \SWGC,\quad \SSWGC.
$
The analysis is carried out in canonical Einstein-frame variables, because this is the frame in which the scalar kinetic term is manifestly canonical and the derivative-based criteria can be stated without ambiguity. The Jordan-frame non-minimal coupling remains part of the physical model through the field redefinition.

A second point is parameter bookkeeping. The genuine shape parameter of the minimal model is $\xi$. The normalization $U_0$ fixes the scalar amplitude once a horizon-crossing point is specified. By contrast, $a$, $b$, and $q$ belong to the FRDSSC and are conjectural inputs rather than model couplings. The SWGC and SSWGC forms adopted below contain no freely adjustable coefficient. This distinction prevents an observational fit from being misidentified as a measurement of a swampland constant.
\begin{table}[t]
\caption{Parameter bookkeeping for the model and for the swampland diagnostics.}
\label{tab:params}
\small
\setlength{\tabcolsep}{4pt}
\renewcommand{\arraystretch}{1.05}
\begin{ruledtabular}
\begin{tabular}{l p{3.6cm} p{1.7cm}}
\textbf{Quantity} & \textbf{Role} & \textbf{Status} \\
\hline
\multicolumn{3}{l}{\textbf{Model parameters}} \\
\hline
$\xi$ & Potential shape & Free parameter \\
$U_0=2\lambda_0$ & Energy scale & Fixed by $A_s$ \\
$x_*=\varphi_*/\mpl$ & Pivot field value & Chosen pivot \\
\hline
\multicolumn{3}{l}{\textbf{FRDSSC parameters}} \\
\hline
$a$ & Slope-Hessian weight & Conjectural input \\
$b=1-a$ & Complementary weight & Derived from $a$ \\
$q$ & FRDSSC exponent ($q>2$) & Conjectural choice \\
\end{tabular}
\end{ruledtabular}
\end{table}

The SWGC and SSWGC expressions used below contain no freely adjustable analogue of $a$ or $b$. This is important: one cannot tune a model coupling to make either scalar weak-gravity inequality true at fixed field once the potential has been specified.
Finally, the observational discussion is deliberately conservative. ACT DR6 combined with Planck and DESI has shifted the preferred scalar tilt upward relative to Planck 2018, with the P-ACT-LB combination giving $n_s=0.9743\pm0.0034$ in the standard analysis, while the addition of BK18 gives the frequently quoted bound $r<0.038$ at 95\% confidence.\cite{3,4} Planck 2018 by itself gives $n_s=0.9649\pm0.0042$ and, with BK15, $r_{0.002}<0.056$ at 95\% confidence.\cite{2} BK18 independently reports $r_{0.05}<0.036$ at 95\% confidence.\cite{BK18}
\section{Einstein-frame formulation of the minimal unimodular model}
\label{sec:model}
For the present purpose it is sufficient to start from the Einstein-frame form obtained after imposing the unimodular constraint and performing the conformal transformation associated with Eq.~\eqref{eq:fjor}. The scalar sector can be written as
\begin{equation}
\begin{split}
&S_E=\int\diff^4x\sqrt{-g}\bigg[
\frac{\mpl^2}{2}R
\\&-\frac12 g^{\mu\nu}\partial_\mu\phi\partial_\nu\phi
-U(\phi)
\bigg].
\label{eq:SE}
\end{split}
\end{equation}
The original derivation obtains a constant Einstein-frame Lagrange multiplier $\lambda_0$ and an effective potential proportional to $2\lambda_0/\Omega^4$.\cite{Malekpour2024} We therefore define
\begin{equation}
U_0\equiv2\lambda_0,
\qquad
D(x)\equiv1+\xi x^2,
\qquad
x\equiv\frac{\varphi}{\mpl},
\end{equation}
so that
\begin{equation}
{U(x)=U_0D^{-2}}.
\label{eq:U}
\end{equation}
The conformal factor is
\begin{equation}
\Omega^2=\frac{2f(\varphi)}{\mpl^2}=1+\frac{\xi\varphi^2}{\mpl^2}=D.
\label{eq:omega}
\end{equation}
The canonical Einstein-frame scalar is related to the Jordan-frame field by
\begin{equation}
\frac{\diff\phi}{\diff\varphi}
=
\frac{\sqrt{1+\xi(1+6\xi)\varphi^2/\mpl^2}}
{1+\xi\varphi^2/\mpl^2}.
\label{eq:canonicalmap}
\end{equation}
It is convenient to define
\begin{equation}
A(x)\equiv1+\xi(1+6\xi)x^2.
\label{eq:A}
\end{equation}
Then
\begin{equation}
\frac{\diff\phi}{\mpl}=\frac{\sqrt{A}}{D}\diff x,
\qquad
\frac{\diff}{\diff(\phi/\mpl)}
=\frac{D}{\sqrt A}\frac{\diff}{\diff x}.
\label{eq:operator}
\end{equation}
The physical regularity domain must satisfy
\begin{equation}
D>0,
\qquad
A>0,
\label{eq:regularity}
\end{equation}
so that the effective Planck mass and scalar kinetic metric remain well-defined. The phenomenological analysis below focuses primarily on the positive-$\xi$ branch, for which these conditions are automatic for real $x$.

\section{Background equations and inflationary observables}
\label{sec:observables}
For the canonical action \eqref{eq:SE}, the homogeneous equations are
\begin{align}
3\mpl^2H^2&=\frac12\dot\phi^2+U,
\label{eq:fried1}\\
-2\mpl^2\dot H&=\dot\phi^2,
\label{eq:fried2}\\
\ddot\phi+3H\dot\phi+U_{,\phi}&=0.
\label{eq:kg}
\end{align}
In the potential slow-roll regime,
\begin{equation}
3\mpl^2H^2\simeq U,
\qquad
3H\dot\phi\simeq-U_{,\phi}.
\label{eq:sr}
\end{equation}
Using Eq.~\eqref{eq:operator}, one obtains
\begin{equation}
\frac{\mpl U_{,\phi}}{U}=-\frac{4\xi x}{\sqrt A}.
\label{eq:d1exact}
\end{equation}
The first two potential slow-roll parameters are therefore
\begin{equation}
\epsV=\frac{8\xi^2x^2}{A}
\label{eq:epsilon}
\end{equation}
and
\begin{equation}
\etaV=
\frac{4\xi\left(24\xi^3x^4+4\xi^2x^4+3\xi x^2-1\right)}{A^2}.
\label{eq:eta}
\end{equation}
At first order in the potential slow-roll expansion,
\begin{equation}
\ns=1-6\epsV+2\etaV,
\qquad
\rr=16\epsV.
\label{eq:nsr}
\end{equation}
Equivalently,
\begin{equation}
\rr=\frac{128\xi^2x^2}{A}
\label{eq:r}
\end{equation}
and
\begin{equation}
\begin{split}
&\ns=
1-
\frac{48\xi^2x^2}{A}
\\&+
\frac{8\xi\left(24\xi^3x^4+4\xi^2x^4+3\xi x^2-1\right)}{A^2}.
\label{eq:ns}
\end{split}
\end{equation}
The leading scalar running is
\begin{equation}
\alphas\simeq16\epsV\etaV-24\epsV^2-2d_1d_3,
\label{eq:running}
\end{equation}
where the dimensionless canonical derivatives are defined by
\begin{equation}
d_n\equiv\mpl^n\frac{U^{(n)}(\phi)}{U}.
\label{eq:dn}
\end{equation}
The normalization is determined by the scalar amplitude,
\begin{equation}
\As\simeq\frac{U_*}{24\pi^2\mpl^4\epsilon_{V*}},
\label{eq:As}
\end{equation}
so that
\begin{equation}
U_0=24\pi^2\As\,\epsilon_{V*}D_*^2\mpl^4.
\label{eq:U0norm}
\end{equation}
This normalization does not change any SWGC or SSWGC conclusion because their dimensionless forms depend only on ratios of derivatives of $U$.
\subsection{Small-field expansion}
\label{subsec:small}
For $|\xi|x^2\ll1$, the exact expressions give
\begin{align}
\epsV&=8\xi^2x^2+\mathcal O(x^4),
\label{eq:epssmall}\\
\etaV&=-4\xi+4\xi^2(5+12\xi)x^2+\mathcal O(x^4),
\label{eq:etasmall}\\
\rr&=128\xi^2x^2+\mathcal O(x^4),
\label{eq:rsmall}\\
\ns&=1-8\xi+8\xi^2(12\xi-1)x^2+\mathcal O(x^4).
\label{eq:nssmall}
\end{align}
At the exact origin,
\begin{equation}
\begin{split}
&\epsV(0)=0,
\quad
\etaV(0)=-4\xi,\\
&\ns(0)=1-8\xi,
\quad
\rr(0)=0.
\label{eq:origin}
\end{split}
\end{equation}
Thus the near-origin estimate for the P-ACT-LB central value is
\begin{equation}
\xi\simeq\frac{1-0.9743}{8}=3.2125\times10^{-3}.
\label{eq:xiactorigin}
\end{equation}
This estimate is only a limiting relation: at $x=0$ the slow-roll parameter $\epsV$ vanishes, so the point itself cannot be used as a literal finite-amplitude pivot prediction.
\subsection{Large-field limit}
\label{subsec:large}
For $x^2\rightarrow\infty$ and $\xi>0$,
\begin{align}
\epsV&\longrightarrow\frac{8\xi}{1+6\xi},
\label{eq:epslarge}\\
\etaV&\longrightarrow\frac{16\xi}{1+6\xi},
\label{eq:etalarge}\\
\rr&\longrightarrow\frac{128\xi}{1+6\xi},
\label{eq:rlarge}\\
\ns&\longrightarrow\frac{1-10\xi}{1+6\xi}.
\label{eq:nslarge}
\end{align}
The asymptotic slow-roll condition $\epsV<1$ is therefore equivalent to $\xi<1/2$. The large-field limit never reproduces the standard Starobinsky prediction $r\rightarrow0$ at fixed positive $\xi$; instead, $r$ approaches a finite value.
\section{E-folds and the end of inflation}
\label{sec:efolds}
The slow-roll number of e-folds is
\begin{equation}
N\simeq\frac{1}{\mpl^2}\int_{\phi_e}^{\phi_*}\frac{U}{U_{,\phi}}\diff\phi.
\label{eq:Ngeneral}
\end{equation}
Using Eqs.~\eqref{eq:operator} and \eqref{eq:U},
\begin{equation}
\frac{U}{U_{,\phi}}\frac{\diff\phi}{\mpl^2}
=-\frac{A}{4\xi xD}\diff x,
\end{equation}
which integrates to
\begin{equation}
N=
\frac{1}{4\xi}\ln\left(\frac{x_e}{x_*}\right)
+\frac34\ln\left(\frac{1+\xi x_e^2}{1+\xi x_*^2}\right).
\label{eq:Nexact}
\end{equation}
The second logarithm is essential. It originates from the canonical field Jacobian and is absent only in an incorrect treatment that replaces $\diff\phi$ by $\mpl\diff x$.
If the end of inflation is assumed to be determined solely by $\epsV=1$, then
\begin{equation}
1=\frac{8\xi^2x_e^2}{1+\xi(1+6\xi)x_e^2},
\end{equation}
so
\begin{equation}
x_e^2=\frac{1}{\xi(2\xi-1)}.
\label{eq:xe}
\end{equation}
A real end point therefore requires
\begin{equation}
\xi>\frac12.
\label{eq:xiexit}
\end{equation}
For $0<\xi\le1/2$ the asymptotic value of $\epsV$ remains below unity and inflation does not terminate by the standard $\epsV=1$ condition.
The same model has, however,
\begin{equation}
\etaV\simeq-4\xi
\end{equation}
near the origin. Consequently conventional slow roll requires $\xi\ll1$. The conditions
\begin{equation}
\xi\ll1
\qquad\text{and}\qquad
\xi>\frac12
\end{equation}
are incompatible. The minimal model therefore does not possess a self-contained near-origin slow-roll phase followed by an $\epsV=1$ exit. This is an analytic structural result and does not depend on a particular benchmark.
A formal large-$N$ solution can be produced by taking $x_*$ very small when $\xi>1/2$, because the logarithm in Eq.~\eqref{eq:Nexact} can become large. Such a configuration is not a controlled slow-roll solution: the same limit gives $|\etaV|\simeq4\xi>2$. The apparent enhancement of $N$ therefore does not rescue the model.
\begin{figure*}[t]
\centering
\includegraphics[width=\linewidth]{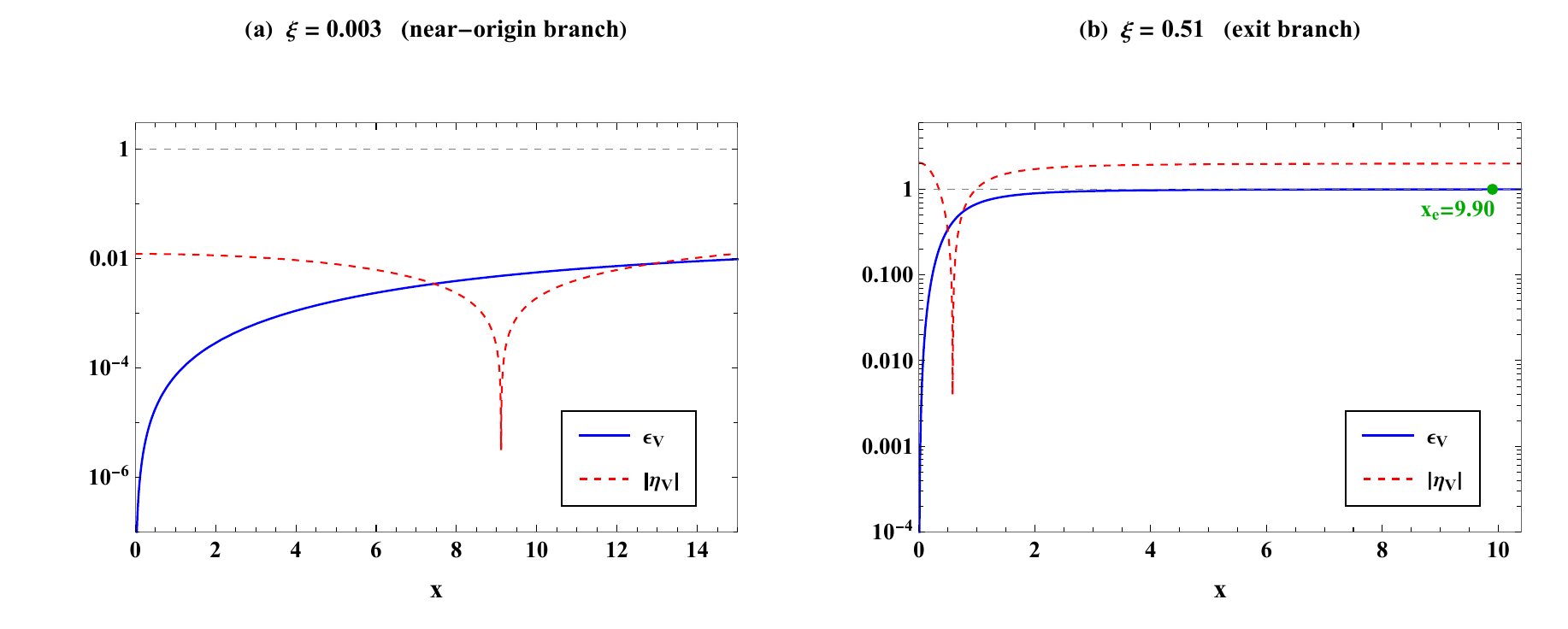}
\caption{Slow-roll parameters $\epsilon_V(x)$ (solid blue) and
$|\eta_V(x)|$ (dashed red), on a logarithmic scale, for (a) the
near-origin branch, $\xi=0.003$, and (b) the exit branch,
$\xi=0.51$. In (a), neither parameter reaches unity over the
plotted range, so no standard exit occurs; the cusp in $|\eta_V|$
near $x\simeq9$ marks a sign change, not an artifact. In (b),
$\epsilon_V$ reaches unity at the analytic exit point
$x_e\simeq9.90$ [Eq.~\eqref{eq:xe}] (green marker), but $|\eta_V|$
has already exceeded unity in its neighborhood.}
\label{fig:exit-tension}
\end{figure*}
Figure~\ref{fig:exit-tension} makes the structural tension of
Sec.~\ref{sec:efolds} explicit. In the near-origin branch
[panel~(a)], the potential slow-roll parameters remain small over a
wide field range, consistent with the observationally preferred
regime $\xi\ll1$; however, $\epsilon_V$ never approaches unity, so
that no standard exit from inflation occurs along this branch. In
the exit branch [panel~(b)], obtained for $\xi>1/2$, the model does
reach $\epsilon_V=1$ at the analytic point $x_e$, but only at the
cost of driving $|\eta_V|$ above unity well before $x_e$ is
attained. The two panels therefore illustrate, at the level of a
single representative parameter choice in each regime, the more
general statement proved in Sec.~\ref{sec:efolds}: the conditions
$\xi\ll1$ (required for a slowly rolling near-origin phase) and
$\xi>1/2$ (required for a finite exit point) are mutually
exclusive, so that the minimal non-minimal unimodular model cannot
simultaneously support a consistent slow-roll phase and a standard
graceful exit.

\subsection{A stronger check of the exit problem}
\label{subsec:exitwindow}
It is useful to check whether the incompatibility found above could be evaded by moving away from the origin. For $\xi>1/2$, the end point $x_e$ exists. However, the slow-roll domain is simultaneously restricted by $\epsV<1$ and $|\etaV|<1$. The lower boundary of this domain is determined by $\etaV=-1$ because $\etaV(0)=-4\xi<-2$. Numerically approaching the critical value $\xi\to1/2^+$ enlarges the available field interval and the formal number of e-folds, but the slow-roll observables remain far from the CMB region. In particular, as $\xi\to1/2^+$, the point where $\etaV=-1$ approaches $x\simeq0.3445$, at which
\begin{equation}
\epsV\simeq0.1919,
\qquad
n_s\simeq-2.15,
\qquad
r\simeq3.07.
\end{equation}
Thus increasing the formal e-fold count by choosing $\xi$ extremely close to $1/2$ does not produce an observationally acceptable slow-roll trajectory. Over the whole positive-$\xi$ exit branch, the portion satisfying $\epsV<1$ and $|\etaV|<1$ has no overlap with the ACT/P-ACT $(n_s,r)$ region. The obstruction is therefore not merely the absence of an exit for small $\xi$; it is the incompatibility of the standard exit branch with the observed slow-roll observables.

\section{The three swampland diagnostics}
\label{sec:swampland}
The canonical Einstein-frame formulation is essential here. The non-minimal Jordan-frame field $\varphi$ is not canonical, so directly differentiating $U(\varphi)$ with respect to $\varphi$ would produce frame-dependent and normalization-dependent statements. The operator in Eq.~\eqref{eq:operator} removes this ambiguity.
\subsection{Further-refined de Sitter conjecture}
\label{subsec:frdssc}
The further-refined de Sitter criterion considered here is\cite{AndriotRoupec}
\begin{equation}
\begin{split}
&\left(\mpl\frac{|U_{,\phi}|}{U}\right)^q
-a\mpl^2\frac{U_{,\phi\phi}}{U}
\ge b,\\
&a>0,
\quad b>0,
\quad a+b=1,
\quad q>2.
\label{eq:frdssc}
\end{split}
\end{equation}
For the present model,
\begin{equation}
\left(\frac{4\xi|x|}{\sqrt A}\right)^q-a\etaV\ge1-a.
\label{eq:frdsscx}
\end{equation}
It is useful to stress the parameter roles. The pair $(\xi,U_0)$ belongs to the model, whereas $(a,b,q)$ labels the conjectural criterion. No CMB observation measures $a$ or $b$ directly.
If $1-\etaV>0$, the inequality can be solved as
\begin{equation}
a\ge a_{\min}(x,\xi;q)
=\frac{1-|d_1|^q}{1-d_2},
\hspace{2pt}
b\le1-a_{\min}.
\label{eq:amin}
\end{equation}
At first order in slow roll,
\begin{equation}
d_1=\sqrt{\frac r8},
\quad
 d_2=\etaV
=\frac12\left(n_s+\frac{3r}{8}-1\right),
\label{eq:obsderiv}
\end{equation}
so
\begin{equation}
a_{\min}(n_s,r;q)=
\frac{1-(r/8)^{q/2}}
{1-\frac12\left(n_s+\frac{3r}{8}-1\right)}.
\label{eq:aminobs}
\end{equation}
In the strict near-origin limit, $d_1\rightarrow0$ and $d_2\rightarrow-4\xi$, yielding
\begin{equation}
a\ge\frac{1}{1+4\xi},
\qquad
b\le\frac{4\xi}{1+4\xi}.
\label{eq:frdsscorigin}
\end{equation}
Combining this with $n_s\simeq1-8\xi$ gives
\begin{equation}
a\ge\frac{2}{3-n_s}
\label{eq:frdssnssmall}
\end{equation}
for the near-origin approximation.
For illustration, the P-ACT-LB central tilt gives
\begin{equation}
a\gtrsim0.9873
\end{equation}
when $r$ is negligible. At a finite-$x$ ACT-centered benchmark used below, the required value is somewhat higher because the nonzero tensor contribution changes both terms in Eq.~\eqref{eq:aminobs}.
\begin{table*}[t]
\centering
\caption{FRDSSC requirements for rectangular observational envelopes at $q=2.2$. These are conservative parameter-space translations, not likelihood-derived measurements of $a$ and $b$.}
\label{tab:frdssc}
\begin{tabular}{l c c c c}
\toprule
\textbf{Dataset} & \textbf{$n_s$ interval} & \textbf{$r$ bound} & \textbf{$\min a_{\min}$} & \textbf{$\max a_{\min}$} \\
\midrule
Planck 2018 ($2\sigma$ tilt) & $0.9565$-$0.9733$ & $0.056$ & $0.9787$ & $0.9929$ \\
P-ACT-LB ($2\sigma$ tilt) & $0.9675$-$0.9811$ & $0.038$ & $0.9840$ & $0.9949$ \\
\bottomrule
\end{tabular}
\end{table*}
The rectangular analysis demonstrates that the FRDSSC can be accommodated, but only by taking $a$ close to one and therefore $b$ at the percent level or below. This is a theoretical compatibility condition, not an empirical determination of a universal constant.

\subsection{Scalar weak gravity conjecture}
\label{subsec:swgc}
The scalar weak gravity criterion adopted here is the
potential form associated with the scalar generalization
of the weak gravity conjecture \cite{Palti,LiuSWGC}.
\begin{equation}
\left(U_{,\phi\phi\phi}\right)^2
\ge
\frac{1}{\mpl^2}\left(U_{,\phi\phi}\right)^2.
\label{eq:swgc}
\end{equation}
In terms of $d_n$,
\begin{equation}
\Delta_{SW}=d_3^2-d_2^2\ge0.
\label{eq:swgcd}
\end{equation}
There is no adjustable conjecture coefficient analogous to $a$ or $b$ in this simple one-field condition.
The third canonical derivative is
\begin{equation}
d_3=-\frac{8x\xi^2}{A^{7/2}}\,P_3(x,\xi),
\label{eq:d3}
\end{equation}
where
\begin{align}
&P_3(x,\xi)={}288\xi^5x^6+96\xi^4x^6+8\xi^3x^6
+54\xi^3x^4\nonumber\\&+9\xi^2x^4
-54\xi^2x^2-6\xi x^2-12\xi-7.
\label{eq:P3}
\end{align}
At $x=0$,
\begin{equation}
d_2=-4\xi,
\qquad d_3=0,
\end{equation}
so
\begin{equation}
\Delta_{SW(0)}=-16\xi^2<0\quad(\xi\ne0).
\label{eq:swgcorigin}
\end{equation}
Thus the SWGC fails exactly at the stationary point for every nonzero $\xi$.
The large-field limit provides an independent diagnostic. One finds
\begin{equation}
d_2\rightarrow\frac{16\xi}{1+6\xi},
\qquad
d_3\rightarrow-\frac{64\xi^{3/2}}{(1+6\xi)^{3/2}},
\end{equation}
and hence
\begin{equation}
\Delta_{SW(\infty)}
=\frac{256\xi^2(10\xi-1)}{(1+6\xi)^3}.
\label{eq:swgclarge}
\end{equation}
The asymptotic SWGC condition is therefore
\begin{equation}
\xi\ge0.1.
\label{eq:swgcinfcond}
\end{equation}
Notice that this lower bound is already much larger than the few-$10^{-3}$ values suggested by the near-origin scalar tilt. The same conclusion is consistent with the direct numerical scan described below.
\begin{figure}[h!]
 \begin{center}
 \includegraphics[height=6cm,width=7cm]{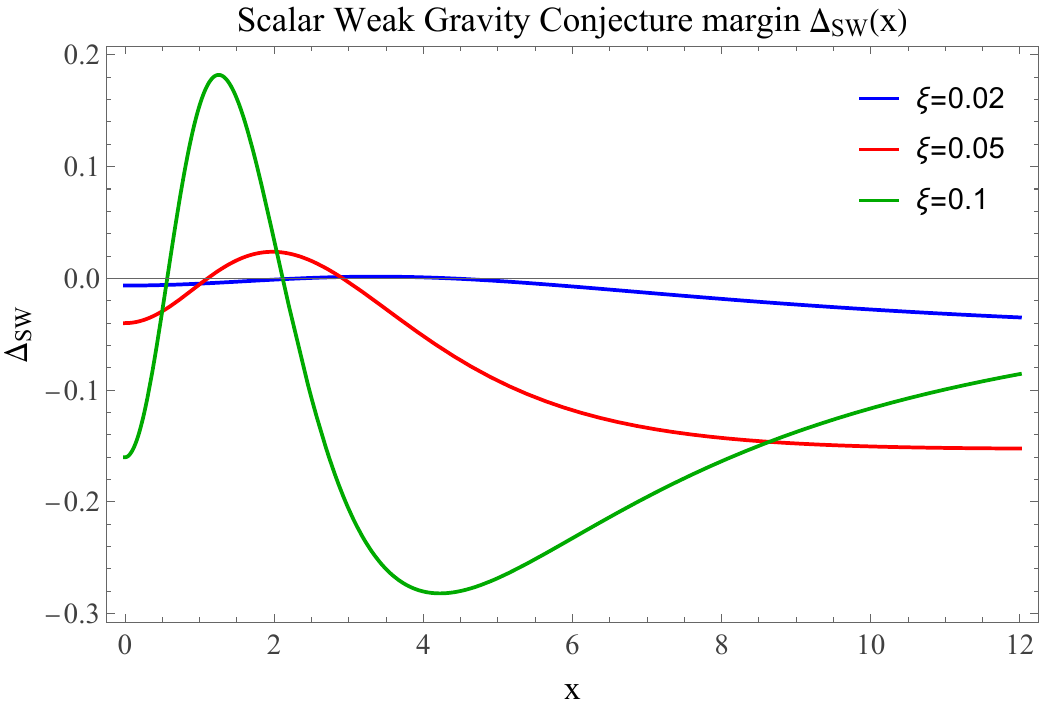}
 \caption{\small{The scalar weak-gravity margin $\Delta_{\rm SW}(x)=d_3^2-d_2^2$ for $\xi=0.02,\,0.05,\,0.1$. The horizontal line marks $\Delta_{\rm SW}=0$; the SWGC holds where the curves lie above it.}}
 \label{fig:swgc-margin}
 \end{center}
 \end{figure}
Fig.~\ref{fig:swgc-margin} shows the SWGC margin $\Delta_{\rm SW}(x)=d_3^2-d_2^2$ as a function of the canonical field for $\xi=0.02,\,0.05,\,0.1$. For each value shown, $\Delta_{\rm SW}$ oscillates between positive and negative values rather than maintaining a fixed sign, so the criterion is satisfied only over finite, disconnected intervals of $x$ rather than throughout the field range. Larger $\xi$ produces a more pronounced oscillation, consistent with the asymptotic threshold $\xi\gtrsim 0.1$ identified in Eq.~(60): near this value the margin briefly turns positive before returning negative at larger $x$, whereas for $\xi=0.02$ the margin remains close to zero and predominantly negative. This behavior illustrates why no single choice of $\xi$ renders the SWGC satisfied uniformly, reinforcing the field-dependent, rather than parameter-dependent, nature of the obstruction discussed in Sec.~VI.B.
\subsection{Strong scalar weak gravity conjecture}
\label{subsec:sswgc}
The strong scalar weak gravity condition is\cite{GonzaloIbanez}
\begin{equation}
2\left(U_{,\phi\phi\phi}\right)^2
-U_{,\phi\phi}U_{,\phi\phi\phi\phi}
\ge
\frac{\left(U_{,\phi\phi}\right)^2}{\mpl^2},
\label{eq:sswgc}
\end{equation}
which becomes
\begin{equation}
\Delta_{SSW}=2d_3^2-d_2d_4-d_2^2\ge0.
\label{eq:sswgcd}
\end{equation}
The conjecture is a field-space statement, and the relevant field is the canonical $\phi$.
The fourth derivative is
\begin{equation}
d_4=\frac{8\xi^2}{A^5}P_4(x,\xi),
\label{eq:d4}
\end{equation}
with the polynomial $P_4$ given in Appendix~\ref{app:d4}. At the origin,
\begin{equation}
d_4(0)=8\xi^2(12\xi+7),
\end{equation}
so
\begin{equation}
\Delta_{SSW(0)}=16\xi^2(24\xi^2+14\xi-1).
\label{eq:sswgcorigin}
\end{equation}
The positive root is
\begin{equation}
\xi_{\rm SSWGC}^{(0)}
=\frac{-14+\sqrt{292}}{48}
\simeq0.0643.
\label{eq:xissorigin}
\end{equation}
Hence the strict near-origin SSWGC and the near-origin ACT tilt are incompatible: $n_s\simeq1-8\xi$ would give $n_s\lesssim0.485$ at the SSWGC threshold, far below the CMB range.
At large field,
\begin{equation}
d_4\rightarrow\frac{256\xi^2}{(1+6\xi)^2},
\end{equation}
and the SSWGC margin simplifies dramatically,
\begin{equation}
\Delta_{SSW(\infty)}
=\frac{256\xi^2(10\xi-1)}{(1+6\xi)^3}.
\label{eq:sswgclarge}
\end{equation}
Thus both scalar weak-gravity criteria have the same asymptotic threshold $\xi=0.1$, although their finite-field behavior differs.
\begin{figure}[h!]
 \begin{center}
 \includegraphics[height=8cm,width=8cm]{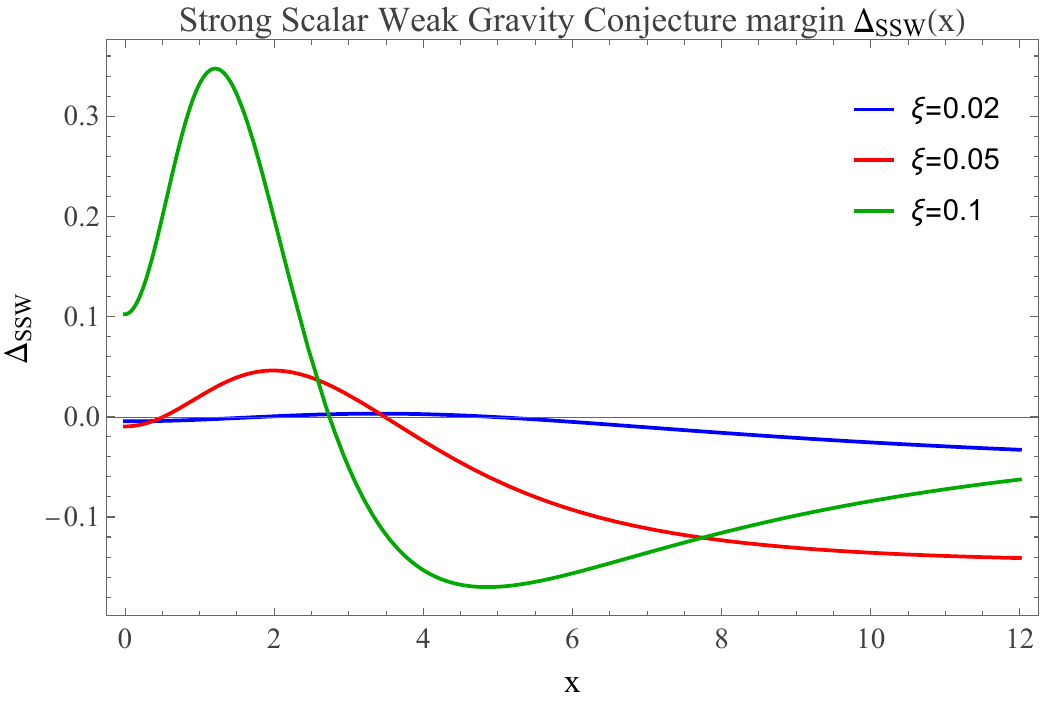}
 \caption{\small{The strong scalar weak-gravity margin $\Delta_{\rm SSW}(x)=2d_3^2-d_2d_4-d_2^2$ for the same $\xi$ values as above. The SSWGC holds where the curves lie above $\Delta_{\rm SSW}=0$.}}
 \label{fig:sswgc-margin}
 \end{center}
 \end{figure}
{Fig.~\ref{fig:sswgc-margin} shows the corresponding SSWGC margin $\Delta_{\rm SSW}(x)=2d_3^2-d_2d_4-d_2^2$ for the same values $\xi=0.02,\,0.05,\,0.1$. The overall pattern mirrors that of $\Delta_{\rm SW}(x)$ in Fig.~\ref{fig:swgc-margin}: each curve alternates in sign rather than remaining positive throughout, and the amplitude of the oscillation grows with $\xi$. The positive excursion near $x\sim 1$-$2$ is somewhat larger here than in the SWGC case, consistent with the SSWGC being the less restrictive of the two conditions, yet it still gives way to a negative region at larger $x$ for every $\xi$ shown. Taken together with Fig.~\ref{fig:swgc-margin}, this indicates that widening the criterion from SWGC to SSWGC does not by itself produce a field range over which both scalar weak-gravity conditions hold uniformly for the considered values of $\xi$.}
\section{Numerical analysis tied to the model parameters}
\label{sec:numerics}
The numerical analysis is performed in the two-dimensional model parameter space $(\xi,x)$ for $\xi>0$. The normalization $U_0$ is then fixed from $A_s$ and does not enter the three dimensionless swampland margins. The conjectural parameters $a$, $b$, and $q$ are varied only when testing the FRDSSC.
The observational envelopes used for a conservative parameter-space scan are
\begin{align}
&0.9565\le n_s\le0.9733,\hspace{3pt}&r<0.056\nonumber\\& \text{(Planck-2018/BK15 reference)},
\label{eq:planckenv}\\
&0.9675\le n_s\le0.9811,\hspace{3pt}&r<0.038\nonumber\\& \text{(P-ACT-LB-BK18 envelope)}.
\label{eq:actenv}
\end{align}
The first is only a rectangular proxy for the Planck scalar constraint and a tensor bound; the second combines the ACT-era scalar tilt with the tensor limit reported for P-ACT-LB-BK18. The full covariance of the likelihoods is not represented by these rectangles, so the scan is a consistency diagnostic rather than a Bayesian parameter inference.
\subsection{ACT-centered finite-field benchmark}
Solving
\begin{equation}
\ns(x_*,\xi)=0.9743
\end{equation}
for $\xi=0.003$ gives
\begin{equation}
\begin{split}
&x_*=5.1413,
\qquad
\epsilon_{V*}=1.7610\times10^{-3},\\
&\eta_{V*}=-7.5669\times10^{-3},
\end{split}
\end{equation}
and therefore
\begin{equation}
n_{s*}=0.9743,
\qquad
r_*=0.02818.
\label{eq:actbenchmark}
\end{equation}
The corresponding swampland margins are
\begin{equation}
\Delta_{SW^*}\simeq-5.28\times10^{-5},
\qquad
\Delta_{SSW^*}\simeq-4.75\times10^{-5}.
\label{eq:actmargins}
\end{equation}
Thus the point is a phenomenological CMB match at the pivot level but fails both scalar weak-gravity diagnostics.
For this point and several values of the FRDSSC exponent,
\begin{table}[h!]
\centering
\caption{FRDSSC requirement at the ACT-centered benchmark, $(\xi,x_*)=(0.003,5.1413)$.}
\label{tab:benchfrd}
\begin{ruledtabular}
\begin{tabular}{c c c}
\textbf{$q$} & \textbf{$a_{\min}$} & \textbf{$b_{\max}=1-a_{\min}$} \\
\hline
$2.01$ & $0.98909$ & $0.01091$ \\
$2.20$ & $0.99050$ & $0.00950$ \\
$2.50$ & $0.99164$ & $0.00836$ \\
$3.00$ & $0.99228$ & $0.00772$ \\
$4.00$ & $0.99248$ & $0.00752$ \\
\end{tabular}
\end{ruledtabular}
\end{table}
The benchmark therefore illustrates a useful separation: the FRDSSC can be accommodated by a conjectural choice of $a$, but the same point cannot satisfy SWGC or SSWGC because those conditions contain no analogous tunable constant.
\subsection{Global scan over observational envelopes}
A numerical scan over
\begin{equation}
10^{-5}\le\xi\le0.5,
\qquad
10^{-5}\le x\le10^3
\end{equation}
with the regular positive-$\xi$ branch retained gives the following maximum margins inside the observational rectangles. The maximization was repeated with continuous optimization after a broad logarithmic pre-scan to reduce sensitivity to grid resolution.
\begin{table*}[t]
\centering
\caption{Maximum scalar weak-gravity margins inside the observational envelopes. A negative maximum means that no point in that envelope satisfies the corresponding conjecture.}
\label{tab:scan}
\begin{ruledtabular}
\begin{tabular}{l c c c c c c}
\textbf{Dataset} & \textbf{$\xi_{\rm max}$} & \textbf{$x_{\rm max}$} & \textbf{$\ns$} & \textbf{$\rr$} & \textbf{$\max\Delta_{\mathrm{SW}}$} & \textbf{$\max\Delta_{\mathrm{SSW}}$} \\
Planck-type envelope & $0.002914$ & $7.798$ & $0.9733$ & $0.0560$ & $-2.78\times10^{-6}$ & $+2.03\times10^{-6}$ \\
P-ACT-LB-BK18 envelope & $0.002072$ & $8.991$ & $0.9811$ & $0.0380$ & $-3.49\times10^{-6}$ & $-1.76\times10^{-6}$ \\
\end{tabular}
\end{ruledtabular}
\end{table*}
The Planck-type envelope contains a very small SSWGC-compatible subregion, but not an SWGC-compatible one. Once the scalar tilt is shifted to the P-ACT-LB region, even the SSWGC maximum becomes negative. This is precisely the kind of distinction that is obscured when one simply reports a checkmark for a selected point.
\begin{figure}[h!]
 \begin{center}
 \includegraphics[height=7cm,width=7cm]{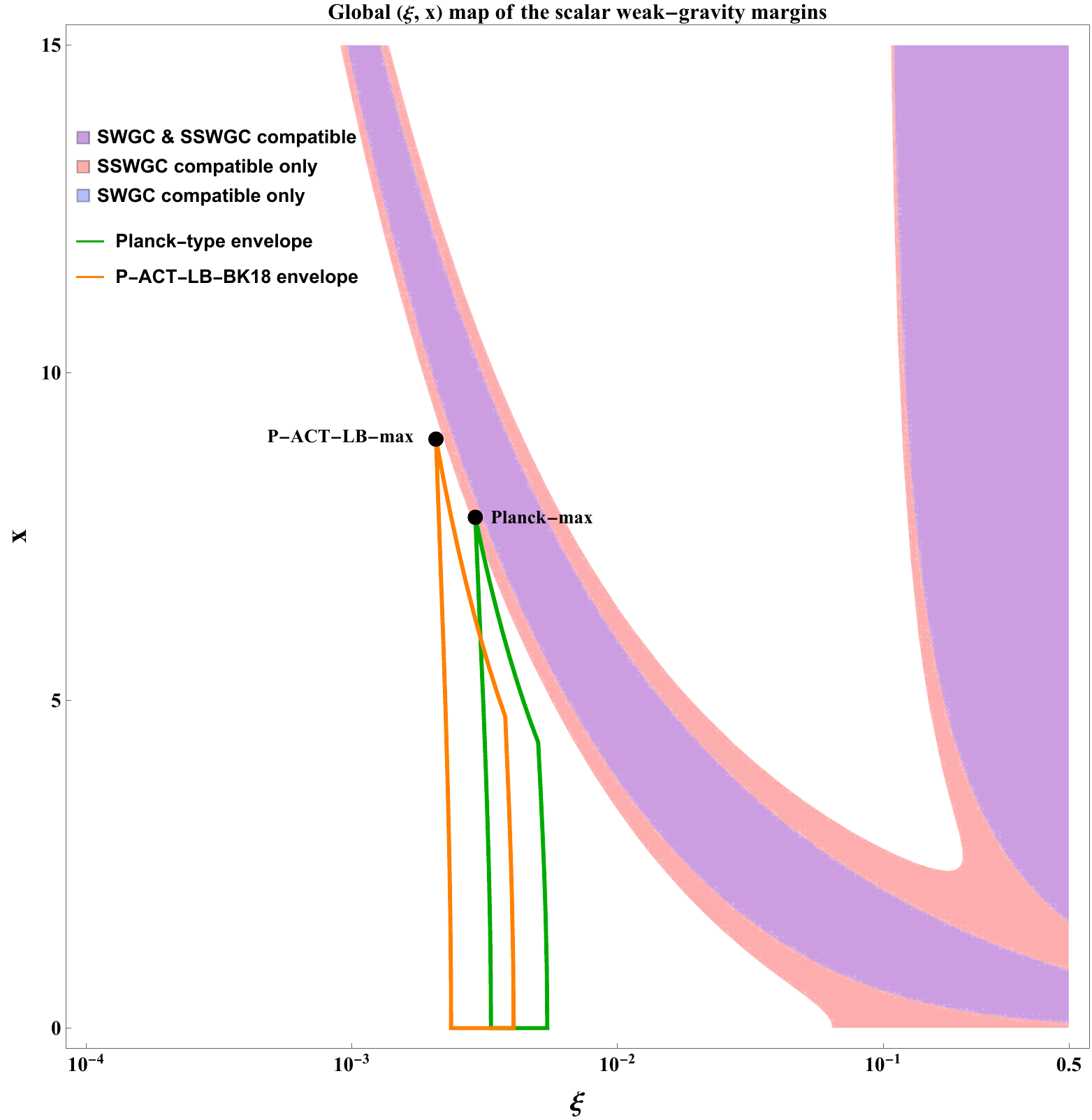}
 \caption{\small{Global $(\xi,x)$ map of the scalar weak-gravity margins. The SWGC- and SSWGC-compatible regions (blue, red) lie entirely outside the Planck-type and P-ACT-LB-BK18 observational envelopes (green, orange), confirming the no-overlap result of Table IV; the marked points give the envelope-constrained maxima of $\Delta_{\rm SW}$ and $\Delta_{\rm SSW}$.}}
 \label{fig:global-map}
 \end{center}
 \end{figure}
Fig.~\ref{fig:global-map} translates this scan into a continuous picture of the $(\xi,x)$ plane. The SWGC- and SSWGC-compatible regions (blue and red) trace a curved band that moves to larger $x$ as $\xi$ decreases, while the Planck-type and P-ACT-LB-BK18 envelopes (green and orange) occupy a narrow strip centered near $\xi\sim 10^{-3}$. The two regions never overlap: the compatible band remains bounded away from both observational envelopes across the entire scanned range, with the closest approach occurring near the envelope-constrained maxima marked on the plot. This confirms, at the level of the full parameter space rather than a single benchmark, that the negative margins reported in the table are not an artifact of the particular grid points chosen but reflect a structural separation between the scalar weak-gravity criteria and the region singled out by current data.
\subsection{Why increasing $\xi$ does not solve the observational problem}
The large-field analysis shows that both scalar weak-gravity criteria can become easier to satisfy for sufficiently large $\xi$, beginning asymptotically at $\xi=0.1$. However, the inflationary observables simultaneously move far away from the CMB region. In particular, the maximal value of $n_s$ for fixed $\xi\ge0.1$ remains well below the observational interval; for example,
\begin{equation}
\max_{x} n_s(x,0.1)\simeq0.2083,
\end{equation}
with larger $\xi$ making the conflict even more severe. Therefore the scalar weak-gravity-compatible regime cannot be connected continuously to the ACT/P-ACT observational region within the minimal potential.
\section{Physical interpretation}
\label{sec:discussion}
The three Swampland criteria considered here provide complementary tests of the same canonically normalized Einstein-frame potential, rather than interchangeable constraints on the model. Their different derivative structures are therefore reflected directly in the regions of parameter space that they probe. The FRDSSC depends on both the first and second derivatives of the potential and can be aided by a negative $U_{,\phi\phi}$. Such a contribution arises naturally in the red-tilted region near the origin, where the curvature of the potential is negative. In addition, the FRDSSC allows a relative weighting of the slope and Hessian contributions through the parameters $a$ and $b=1-a$. A phenomenologically viable point can therefore satisfy this criterion without any modification of the underlying inflationary potential. This should, however, be interpreted as compatibility with the conjectured inequality rather than as evidence for an additional physical sector in the inflationary model.

The SWGC exhibits a qualitatively different behavior because its derivative structure makes it particularly sensitive to stationary points of the potential. At such a point, the cubic derivative vanishes whereas the second derivative remains finite. The corresponding inequality consequently fails for every nonzero $\xi$. This obstruction is intrinsic to the shape of the potential and cannot be removed by changing its overall normalization. In particular, a rescaling of $U_0$ leaves the criterion unchanged because all terms entering the SWGC scale homogeneously with the potential normalization. A second indication of the same tension is obtained in the asymptotic large-field regime: the SWGC becomes positive only for approximately $\xi\geq0.1$, whereas the phenomenologically relevant region associated with the observed scalar tilt lies at much smaller $\xi$. The large-field behavior therefore does not rescue the parameter region preferred by the inflationary observables.

The SSWGC includes the fourth derivative of the potential and can consequently display a different finite-field behavior from the ordinary SWGC. This distinction is visible in the numerical analysis. Within the Planck-compatible region, a small positive SSWGC margin remains possible, demonstrating that the two scalar Swampland conditions do not select exactly the same subset of parameter space. The situation changes, however, when the scalar tilt is shifted to the more recent P-ACT-LB observational envelope: the remaining SSWGC-compatible subregion disappears. The appropriate conclusion is therefore not that the SSWGC is intrinsically weaker than the SWGC, but rather that its finite-field constraint is less restrictive over a limited part of the parameter space that remains compatible with the older Planck-era bounds.

The interpretation of these results also requires careful parameter counting. At the level relevant for the scalar power spectrum, the minimal model is characterized by the two quantities $\xi$ and $U_0$. Once a pivot solution has been selected, the overall normalization $U_0$ is fixed by the measured scalar amplitude, leaving $\xi$ as the principal shape parameter controlling the inflationary predictions. The FRDSSC introduces the additional conjectural quantities $a$, $b$, and $q$, but these parameters should not be interpreted as independent physical couplings of unimodular inflation. By contrast, neither the SWGC nor the SSWGC introduces an analogous adjustable conjecture parameter. It follows that the formal possibility of satisfying the FRDSSC by taking $a\rightarrow1$ should not be presented as an additional observational degree of freedom of the inflationary theory. Rather, it specifies how the conjectured criterion itself is realized for a given potential.

A similar distinction is necessary when interpreting the observational comparison. The ACT DR6 collaboration has released likelihoods and associated data products, while analyses combining ACT with Planck and large-scale-structure information have reported a preferred scalar spectral index higher than the value inferred from Planck 2018 alone \cite{3,4}. At the same time, the inferred value of $n_s$ is not completely independent of the precise combination of CMB and low-redshift datasets used in the analysis. Subsequent studies incorporating newer SPT and DESI information can shift the preferred scalar tilt relative to the original P-ACT-LB result \cite{ACTSPTdiscussion}. We therefore use the ACT-era value adopted in this work as a well-defined benchmark for testing the model, rather than treating it as a unique final determination that supersedes all other cosmological analyses.

The post-inflationary history introduces another important layer of interpretation. The relation between an observational pivot scale and a model variable such as $x_*$ is not determined by the inflationary potential alone; it also depends on the subsequent evolution of the Universe and, in particular, on the reheating history. Recent ACT-era analyses of standard attractor models have shown explicitly that assumptions about reheating can modify the inferred inflationary parameter space \cite{31'}. For the minimal unimodular model studied here, however, the issue arises at a more fundamental level. Since the standard $\epsilon_V=1$ exit is absent in the small-$\xi$ regime of phenomenological interest, it is not self-consistent to impose a conventional $N_*=50$-$60$ inflationary trajectory and subsequently treat that value of $N_*$ as a quantity derived from the model. A consistent extension must first specify how inflation ends and how the reheating phase proceeds. Only after these ingredients are established can a fully controlled inference from the pivot scale to the underlying model parameters be performed.
\section{Relation to non-minimal unimodular inflation}
\label{sec:previous}
The present analysis builds on the non-minimal unimodular construction, in which the non-minimal coupling generates an effective Einstein-frame potential, and viable numerical regions capable of reproducing Planck-era inflationary observables were identified \cite{Malekpour2024}. The main difference here is not the basic origin of the effective potential, but the order in which the various consistency requirements are imposed. We first reconstruct the Einstein-frame theory together with its canonical scalar variable. The exact relation between the canonical field and the original non-canonical variable is then used to derive the e-fold relation. Only after this canonical formulation has been established are the Swampland criteria evaluated. Finally, observationally acceptable pivot points are distinguished explicitly from complete inflationary trajectories.

This ordering is essential for a consistent interpretation of the results. The original formulation contains additional Jordan-frame ingredients, including a power-law ansatz for the unimodular multiplier. Those structures are not reintroduced here as independent parameters because the Einstein-frame potential relevant for the present Swampland analysis has already reduced the minimal construction to the two quantities $U_0$ and $\xi$. Reintroducing the Jordan-frame ansatz at this stage without consistently carrying all of its effects through the complete field and frame transformation would amount to combining different parameterizations of the same construction. Such a procedure could artificially enlarge the apparent parameter space and obscure the distinction between genuine physical freedom and quantities that have already been fixed by the Einstein-frame reduction.

The present formulation therefore provides a controlled way of comparing the phenomenological success of the original construction with its dynamical and Swampland consistency. In particular, a parameter choice that reproduces an acceptable pivot-scale spectrum is not automatically promoted to a complete inflationary solution, and satisfaction of one conjectural criterion is not taken as evidence that the remaining criteria are simultaneously fulfilled. This separation of roles is central to the physical interpretation of the minimal model and to identifying which structural modification would be required for a viable extension.
\section{Conclusion}
\label{sec:conclusion}
We have examined the non-minimal unimodular inflationary model in the Einstein frame, with particular emphasis on its slow-roll dynamics, observational viability, and consistency with the scalar-sector Swampland criteria. The analysis is based on the effective potential
$
U(\varphi)=\frac{U_0}{\left(1+\xi\varphi^2/\mpl^2\right)^2},
$
together with the nontrivial canonical field transformation induced by the original non-minimal coupling. Working directly with the canonically normalized field allows the inflationary observables and the derivative-based Swampland conditions to be evaluated within a single, unambiguous framework.

The exact slow-roll relations reveal an important limitation of the minimal construction. In particular, the region of small $\xi$ that is relevant for obtaining phenomenologically interesting inflationary predictions is incompatible with the conventional end-of-inflation condition $\epsilon_V=1$. This incompatibility is analytic and does not depend on the choice of a particular numerical benchmark. Consequently, although the model can reproduce viable observables at an appropriate pivot scale, it cannot by itself realize a conventional, self-contained slow-roll evolution lasting $50$-$60$ e-folds and terminating through the standard condition $\epsilon_V=1$. A successful realization therefore requires additional dynamics beyond the minimal setup.

The Swampland analysis provides a second, complementary layer of constraints. The FRDSSC can be satisfied for values of the parameter $a$ close to unity, although the allowed interval depends on the location in parameter space and, in particular, on the choice of $q$. This dependence is important for the interpretation of the result: the corresponding values of $a$ and $b$ should be regarded as conjectural parameters specifying the strength of the Swampland criterion, rather than as quantities measured by the cosmological data. The SWGC is more restrictive. For every nonzero $\xi$, it fails at the stationary point and does not survive within the observationally allowed region explored in the positive-$\xi$ scan. The SSWGC behaves somewhat less severely, since a small Planck-compatible region can remain viable; nevertheless, it is violated throughout the P-ACT-LB allowed region for the minimal model considered here.
The distinction between an observationally successful pivot point and a globally viable inflationary history is illustrated by the representative choice
$
\xi=0.003,\hspace{1.5pt} x_*=5.1413,
$
for which
$
n_s=0.9743,\hspace{1.5pt} r=0.02818.
$
Thus, the model can reproduce a representative ACT-era pivot point within the observationally relevant range. However, at this same point the SWGC and SSWGC margins are negative. The example therefore makes clear that fitting the scalar and tensor observables at a single pivot scale does not guarantee consistency with the scalar Swampland criteria, nor does it resolve the independent problem associated with completing the inflationary evolution.
The overall result is most naturally understood as a controlled tension among three distinct requirements, observational pivot fit, dynamically complete inflation, scalar swampland consistency. Within the minimal model studied here, these conditions cannot be realized simultaneously. The obstruction is therefore not simply a failure of phenomenological fitting, but a mismatch between the local success of the model at the observational pivot and its global dynamical and theoretical consistency.

It is important to emphasize the scope of this conclusion. The results do not imply that unimodular gravity as a framework is incompatible with quantum gravity, nor do they exclude non-minimal unimodular inflationary models in general. Rather, they identify a specific obstruction associated with the minimal potential and its canonical scalar dynamics. A viable extension would consequently need to modify at least one of the structures responsible for this tension, such as the mechanism governing the end of inflation, the functional form of the scalar potential, or the hierarchy of its canonical derivatives entering the Swampland tests. Any such extension should then be assessed within the same Einstein-frame, canonically normalized formulation, while treating the observational pivot, the complete e-fold history, reheating dynamics, and Swampland conditions as distinct requirements that must ultimately be satisfied in a mutually consistent manner.
\appendix
\section{Canonical derivative algebra}
\label{app:derivatives}
This appendix sets up the algebra needed for taking derivatives with respect to the canonically normalized Einstein-frame field. In particular, we introduce the operator
$
\mathcal{D}_\phi \equiv \frac{D}{\sqrt{A}}\,\frac{\mathrm{d}}{\mathrm{d}x},
$
and use it to construct the first four normalized derivatives,
$
d_n=\frac{\mathcal{D}_\phi^n u}{u}.
$
Their exact forms are written explicitly in terms of \(x\) and \(\xi\). These expressions provide the derivative hierarchy used throughout the subsequent analysis of the Swampland conditions.
Define
\begin{equation}
\mathcal D_\phi\equiv\frac{\diff}{\diff(\phi/\mpl)}
=\frac{D}{\sqrt A}\frac{\diff}{\diff x},
\qquad
u(x)=D^{-2}.
\end{equation}
Then
\begin{equation}
d_n=\frac{1}{\nu}\mathcal D_\phi^n\nu.
\end{equation}
The first four derivatives are
\begin{align}
d_1&=-\frac{4x\xi}{\sqrt A},
\\
d_2&=\frac{4\xi(24\xi^3x^4+4\xi^2x^4+3\xi x^2-1)}{A^2},
\\
d_3&=-\frac{8x\xi^2}{A^{7/2}}P_3(x,\xi),
\\
d_4&=\frac{8\xi^2}{A^5}P_4(x,\xi).
\end{align}
\section{Fourth derivative polynomial}
\label{app:d4}
Here we give the full expression for the polynomial \(P_4(x,\xi)\) entering the fourth canonical derivative \(d_4\). Although this polynomial is lengthy, keeping its exact form is useful because the fourth derivative enters directly into the strong scalar weak gravity conjecture (SSWGC). The same expression is also needed when deriving the analytic behavior of the model in the large-field limit.
The polynomial entering $d_4$ is
\begin{equation}
\begin{split}
&P_4(x,\xi)={}6912\xi^8x^{10}+3456\xi^7x^{10}+576\xi^6x^{10}\\&+32\xi^5x^{10}
+1080\xi^6x^8+360\xi^5x^8+30\xi^4x^8\\
&-3960\xi^5x^6-1230\xi^4x^6-95\xi^3x^6
\\
&-2016\xi^4x^4-1224\xi^3x^4-145\xi^2x^4
\\
&-432\xi^3x^2-198\xi^2x^2-45\xi x^2+12\xi+7.
\label{eq:P4}
\end{split}
\end{equation}
\section{Small-field checks and observational target matching}
\label{app:smallchecks}
This appendix examines the theory close to the origin, where the slow-roll quantities admit simple series expansions. In this limit, the scalar spectral index reduces to the familiar leading-order relation
$
n_s \simeq 1-8\xi.
$
We then indicate which values of \(\xi\) correspond to representative observational values of \(n_s\). For a finite-field benchmark centered on the ACT result, the scalar-amplitude condition is also used to determine the overall normalization \(U_0\). This makes explicit that \(U_0\) fixes the energy scale of the potential, while the Swampland diagnostics considered here depend only on its shape and are therefore unchanged by this normalization.
Expanding the exact expressions gives
\begin{align}
\epsV&=8\xi^2x^2-8\xi^3(1+6\xi)x^4+\mathcal O(x^6),
\\
\etaV&=-4\xi+4\xi^2(5+12\xi)x^2+\mathcal O(x^4),
\\
\ns&=1-8\xi+8\xi^2(12\xi-1)x^2+\mathcal O(x^4).
\end{align}
The origin relation $n_s=1-8\xi$ gives
\begin{align}
\xi&=4.3875\times10^{-3} &&\text{for }n_s=0.9650,\\
\xi&=3.2125\times10^{-3} &&\text{for }n_s=0.9743.
\end{align}
These are limiting values, not complete horizon-crossing solutions.
For the finite-field ACT-centered benchmark,
\begin{equation}
(\xi,x_*)=(0.003,5.1413249)
\end{equation}
solves $n_s=0.9743$ and yields
\begin{equation}
r=0.0281765.
\end{equation}
The scalar amplitude normalization with $A_s=2.1\times10^{-9}$ gives
\begin{equation}
\frac{U_0}{\mpl^4}\simeq1.02\times10^{-9},
\qquad
\frac{U_0^{1/4}}{\mpl}\simeq5.65\times10^{-3}.
\end{equation}
These normalization values are included only to demonstrate that the scalar amplitude can fix the overall scale; they do not alter the shape-based swampland conclusions.
\section{Analytic large-field weak-gravity limits}
\label{app:largechecks}
The large-field regime is treated analytically by taking the asymptotic forms of the canonical derivatives \(d_2\), \(d_3\), and \(d_4\). Substitution into the SWGC and SSWGC expressions gives their corresponding asymptotic margins. Both criteria lead to the same limiting condition,
$
\xi=0.1,
$
as the boundary of the asymptotically allowed region. The appendix also compares this regime with the inflationary observables. In particular, the resulting tensor-to-scalar ratio is incompatible with the observational tensor bound, showing that the large-field region that helps the scalar weak-gravity conditions cannot at the same time provide a phenomenologically acceptable inflationary regime.
Using the exact derivative operator,
\begin{align}
d_2&\rightarrow\frac{16\xi}{1+6\xi},\\
d_3&\rightarrow-\frac{64\xi^{3/2}}{(1+6\xi)^{3/2}},\\
d_4&\rightarrow\frac{256\xi^2}{(1+6\xi)^2}.
\end{align}
Therefore
\begin{align}
\Delta_{SW(\infty)}
&=\frac{256\xi^2(10\xi-1)}{(1+6\xi)^3},\\
\Delta_{SSW(\infty)}
&=\frac{256\xi^2(10\xi-1)}{(1+6\xi)^3}.
\end{align}
The two scalar criteria therefore share the asymptotic boundary $\xi=0.1$.
For $\xi\ge0.1$, the large-field tensor prediction obeys
\begin{equation}
r_\infty=\frac{128\xi}{1+6\xi}\ge8,
\end{equation}
which is already far outside any inflationary tensor bound. This reinforces, independently of the scalar tilt, that the asymptotic scalar-weak-gravity regime cannot be used as an observational inflationary regime of the minimal model.
\section{FRDSSC limits at representative observational points}
\label{app:frlimits}
This appendix focuses on the further-refined de Sitter conjecture at several representative observational locations. For both the Planck and P-ACT-LB rectangular envelopes, as well as for the ACT-centered benchmark, we determine the minimum value of the conjectural parameter \(a\) required by the FRDSSC and quote the corresponding upper value of \(b=1-a\). These results show that satisfying the FRDSSC within the observationally relevant region requires \(a\) to lie close to unity. The resulting numbers are therefore best understood as choices within the conjectural criterion itself, rather than as experimentally determined constants of the inflationary model.
For $q=2.2$, the rectangular envelopes give
\begin{align}
\text{Planck:}\quad&0.9787\lesssim a_{\min}\lesssim0.9929,\\
\text{P-ACT-LB:}\quad&0.9840\lesssim a_{\min}\lesssim0.9949.
\end{align}
For the ACT-centered benchmark,
\begin{equation}
a_{\min}\simeq0.99050,
\qquad
b_{\max}\simeq0.00950.
\end{equation}
The fact that these numbers lie close to the upper end of the allowed $a$ range is a property of the red-tilted potential, not an independently measured value of a swampland constant.\\
\section{Parameter-space consistency logic}
\label{app:logic}
The final appendix collects the main logical relations among the different ingredients of the analysis. The physical parameters of the minimal model,
$
(\xi,U_0),
$
determine the inflationary predictions, including
$
(n_s,r,\alpha_s),
$
whereas the quantities
$
(a,b,q)
$
enter only through the FRDSSC and should not be regarded as additional couplings of the underlying inflationary theory. The SWGC and SSWGC are likewise treated as separate consistency tests rather than as fitted model parameters. This distinction is important for interpreting the results: reproducing the observables at a particular pivot point does not by itself establish a complete \(50\)-\(60\) e-fold inflationary trajectory, and agreement with one Swampland criterion does not imply that the remaining criteria are satisfied or that an ultraviolet completion has been demonstrated.
The minimal model contains two quantities that enter the scalar spectrum,
\begin{equation}
(\xi,U_0).
\end{equation}
The scalar amplitude fixes $U_0$ once $(\xi,x_*)$ is selected. The remaining shape information is controlled by $(\xi,x_*)$. The FRDSSC adds the conjectural labels $(a,b,q)$, with $b=1-a$. SWGC and SSWGC contain no adjustable coefficient in the forms used here.
The logical hierarchy is therefore
\begin{equation}
(\xi,x_*)
\longrightarrow
(n_s,r,\alpha_s),
\end{equation}
followed by the independent tests
\begin{equation}
\FRDSSC(a,q),
\qquad
\SWGC,
\qquad
\SSWGC.
\end{equation}
A point in the $(n_s,r)$ plane cannot be promoted to a viable inflationary solution unless a compatible $N_*$ and end-of-inflation prescription also exist. Likewise, passing one conjecture cannot be promoted to a statement about ultraviolet completion.
\section*{Acknowledgments}
AI-assisted writing statement. The authors employed AI-based utilities during the drafting phase solely to polish the language, rectify grammatical errors, and facilitate \LaTeX\ typesetting. Subsequent to utilizing these resources, the authors carefully reviewed all sections, made necessary modifications, and bear sole responsibility for the content of the submitted article. It is hereby declared that every analytical outcome, conceptual framework, and interpretative assertion originates from the authors' independent scholarly work.


\begin{thebibliography}{99}

\bibitem{2} Planck Collaboration, Y. Akrami et al., Planck 2018 results. X. Constraints on inflation, Astron. Astrophys. 641 (2020) A10, arXiv:1807.06211 [astro-ph.CO].

\bibitem{3}
T. Louis et al. [ACT Collaboration],
The Atacama Cosmology Telescope: DR6 power spectra, likelihoods and
$\Lambda$CDM parameters,
JCAP 11 (2025) 062,
arXiv:2503.14452 [astro-ph.CO].

\bibitem{4}
E. Calabrese et al. [ACT Collaboration],
The Atacama Cosmology Telescope: DR6 constraints on extended
cosmological models,
JCAP 11 (2025) 063,
arXiv:2503.14454 [astro-ph.CO].


\bibitem{6} S. Aoki, H. Otsuka, and R. Yanagita, Higgs-Modular Inflation, arXiv:2504.01622 [hep-ph].

\bibitem{7}
S. D. Odintsov and V. K. Oikonomou,
GW170817 viable Einstein-Gauss-Bonnet inflation compatible with
the Atacama Cosmology Telescope data,
Phys. Lett. B 868 (2025) 139779,
arXiv:2506.08193 [gr-qc].

\bibitem{8}
M. R. Haque and D. Maity,
Minimal plateau inflation in light of ACT DR6 observations,
Phys. Lett. B 873 (2026) 140187,
arXiv:2505.18267 [astro-ph.CO],
doi:10.1016/j.physletb.2026.140187.

\bibitem{9} C. Dioguardi, A. J. Iovino, and A. Racioppi, Fractional attractors in light of the latest ACT observations, Phys. Lett. B 868 (2025) 139664, arXiv:2504.02809 [gr-qc].

\bibitem{10}
A. Salvio,
Independent connection in action during inflation,
Phys. Rev. D 112, L061301 (2025),
arXiv:2504.10488 [hep-ph],
doi:10.1103/tq3v-vy3y.

\bibitem{15}
M. R. Haque, S. Pal, and D. Paul,
ACT DR6 insights on $\alpha$-attractor inflationary models
and reheating,
Phys. Rev. D 114, 023521 (2026),
arXiv:2505.01517 [astro-ph.CO],
doi:10.1103/1kj4-s4ts.

\bibitem{16} N. Bernal and F. Hajkarim, Primordial Gravitational Waves in Nonstandard Cosmologies, Phys. Rev. D 100 no. 6, (2019) 063502, arXiv:1905.10410 [astro-ph.CO].

\bibitem{17} M. R. Haque, D. Maity, T. Paul, and L. Sriramkumar, Decoding the phases of early and late time reheating through imprints on primordial gravitational waves, Phys. Rev. D 104 no. 6, (2021) 063513, arXiv:2105.09242 [astro-ph.CO].

\bibitem{21} K. Bamba, E. Elizalde, S. D. Odintsov, and T. Paul, Inflationary magnetogenesis with reheating phase from higher curvature coupling, JCAP 04 (2021) 009, arXiv:2012.12742 [gr-qc].

\bibitem{22} D. Maity, S. Pal, and T. Paul, Effective Theory of Inflationary Magnetogenesis and Constraints on Reheating, JCAP 05 (2021) 045, arXiv:2103.02411 [hep-th].

\bibitem{23}
M. R. Haque, S. Pal, and D. Paul,
Improved predictions on Higgs-Starobinsky inflation and
reheating with ACT DR6 and primordial gravitational waves,
Phys. Lett. B 869, 139852 (2025),
arXiv:2505.04615 [astro-ph.CO],
doi:10.1016/j.physletb.2025.139852.


\bibitem{k}
N. Schöneberg et al.,
News from the swampland---constraining string theory with
astrophysics and cosmology,
JCAP 10 (2023) 039,
doi:10.1088/1475-7516/2023/10/039.

\bibitem{o}
D. Harlow, B. Heidenreich, M. Reece, and T. Rudelius,
Weak gravity conjecture,
Rev. Mod. Phys. 95, 035003 (2023),
doi:10.1103/RevModPhys.95.035003.

\bibitem{bbb}
M. van Beest, J. Calderón-Infante, D. Mirfendereski,
and I. Valenzuela,
Lectures on the swampland program in string compactifications,
Phys. Rep. 989, 1--50 (2022),
arXiv:2102.01111 [hep-th],
doi:10.1016/j.physrep.2022.09.002.

\bibitem{50}
Achúcarro, Ana, and Gonzalo A. Palma. "The string swampland constraints require multi-field inflation." Journal of Cosmology and Astroparticle Physics 2019.02 (2019): 041-041.

\bibitem{51}
Kehagias, Alex, and Antonio Riotto. A note on Inflation and the Swampland. Fortschritte der Physik 66.10 (2018): 1800052.
\bibitem{52}
Das, Suratna. Warm Inflation in the light of Swampland Criteria. Physical Review D 99.6 (2019): 063514.
\bibitem{53}
S. N. Gashti et al.,
Eternal inflation under swampland constraints: Narrow paths to a
multiverse,
Int. J. Geom. Methods Mod. Phys. (2026),
doi:10.1142/S0219887826503895.

\bibitem{54}
Das, Suratna, Gaurav Goswami, and Chethan Krishnan. Swampland, axions, and minimal warm inflation. Physical Review D 101.10 (2020): 103529.
\bibitem{55}
Motaharfar, Meysam, Vahid Kamali, and Rudnei O. Ramos. Warm inflation as a way out of the swampland. Physical Review D 99.6 (2019): 063513.
\bibitem{56}
Matsui, Hiroki, and Fuminobu Takahashi. Eternal inflation and swampland conjectures. Physical Review D 99.2 (2019): 023533.
\bibitem{57}
Gashti, Saeed Noori, et al. From theory to observation: Swampland conjectures and ACT constraints on marginally deformed R2 inflation. Physics Letters B 876 (2026): 140398.
\bibitem{58}
Lehnert, Kay. Hitchhiker's Guide to the Swampland: The Cosmologist's Handbook to the String‐Theoretical Swampland Programme. Fortschritte der Physik 74.4 (2026): e70073.
\bibitem{59}
S. N. Gashti,
Inflationary dynamics in modified gravity: Implications from
swampland/landscape criteria and P-ACT-LBBK18 observational data,
Phys. Lett. B 880 (2026) 140797,
doi:10.1016/j.physletb.2026.140797.

\bibitem{60}
Taghavi, Sara, Tayeb Golanbari, and Khaled Saaidi. Tsallis holographic inflation in f (R, T) gravity: Planck/ACT constraints, reheating dynamics, and swampland bounds. Progress of Theoretical and Experimental Physics 2026.5 (2026): 053E01.
\bibitem{61}
Yuennan, Jureeporn, Phongpichit Channuie, and Davood Momeni. Warm non-minimally coupled Peccei–Quinn inflation and de Sitter Swampland conjecture. Nuclear Physics B 1012 (2025): 116810.
\bibitem{62}
Sadiq, Shama, et al. D-brane and exponential potentials inspire warm inflation and swampland conjecture. Nuclear Physics B 1018 (2025): 116981.

\bibitem{18'} P. G. Ferreira et al., Scale-independent $R^2$ inflation, Phys. Rev. D 100.12 (2019) 123516.

\bibitem{19'}
D. M. Ghilencea,
Weyl $R^2$ inflation with an emergent Planck scale,
JHEP 10 (2019) 209,
arXiv:1906.11572 [gr-qc],
doi:10.1007/JHEP10(2019)209.

\bibitem{20'} Y. Tang and Y.-L. Wu, Weyl scaling invariant $R^2$ gravity for inflation and dark matter, Phys. Lett. B 809 (2020) 135716.

\bibitem{21'} Q.-Y. Wang, Inflation in light of ACT/SPT: a new perspective from Weyl gravity, arXiv:2512.10862 (2025).


\bibitem{22'}
S. Aoki, H. Otsuka, and R. Yanagita,
Heavy field effects on inflationary models in light of ACT data,
JCAP 11 (2025) 088,
doi:10.1088/1475-7516/2025/11/088.

\bibitem{23'} Z.-Z. Peng, Z.-C. Chen, and L. Liu, The polynomial potential inflation in light of ACT observations, arXiv:2505.12816 (2025).

\bibitem{24'}
I. D. Gialamas, A. Karam, A. Racioppi, and M. Raidal,
Has ACT measured radiative corrections to the tree-level
Higgs-like inflation?,
Phys. Rev. D 112, 103544 (2025),
arXiv:2504.06002 [astro-ph.CO],
doi:10.1103/6fpc-67s1.

\bibitem{25'}
V. K. Oikonomou,
Strong gravity effects on $R^2$-corrected single scalar field
inflation and compatibility with the ACT data,
Phys. Lett. B 871, 139972 (2025),
doi:10.1016/j.physletb.2025.139972.

\bibitem{26'}
V. K. Oikonomou,
String corrected scalar field inflation compatible with the ACT data,
Nucl. Phys. B 1026, 117437 (2026),
arXiv:2508.19196.


\bibitem{27'}
S. V. Ketov, E. O. Pozdeeva, and S. Yu. Vernov,
Inflation in $F(R)$ gravity models revisited after ACT,
JCAP 12 (2025) 040,
arXiv:2508.08927 [gr-qc],
doi:10.1088/1475-7516/2025/12/040.

\bibitem{28'}
S. D. Odintsov and T. Paul,
ACT inflation and its influence on reheating era in
Einstein-Gauss-Bonnet gravity,
Phys. Lett. B 870 (2025) 139930,
doi:10.1016/j.physletb.2025.139930.

\bibitem{29'} W. Ahmed and M. U. Rehman, Radiatively corrected Starobinsky inflation and primordial gravitational waves in light of ACT observations, Phys. Rev. D 112.6 (2025) 063519.

\bibitem{30'}
A. Addazi, Y. Aldabergenov, and S. V. Ketov,
Curvature corrections to Starobinsky inflation can explain the
ACT results,
Phys. Lett. B 869, 139883 (2025),
arXiv:2505.10305,
doi:10.1016/j.physletb.2025.139883.

\bibitem{31'} D. S. Zharov, O. O. Sobol, and S. I. Vilchinskii, ACT observations, reheating, and Starobinsky and Higgs inflation, Phys. Rev. D 112.2 (2025) 023544.


\bibitem{32'}
J. Yuennan, F. Atamurotov, and P. Channuie,
ACT constraints on marginally deformed Starobinsky inflation,
Phys. Lett. B 872, 140065 (2026),
arXiv:2509.23329 [gr-qc],
doi:10.1016/j.physletb.2025.140065.


\bibitem{33'}
F.-Y. Zhang, R. Zhai, and L.-Y. Chen,
Reconciling the ACT preference in $f(T)$ gravity: Inflation and
reheating constraints,
Chin. Phys. C 50 (2026) 065107,
arXiv:2508.16530 [gr-qc],
doi:10.1088/1674-1137/ae5806.


\bibitem{34'}
A. Mohammadi, Yogesh, and A. Wang,
Power law plateau inflation and primordial gravitational waves
in the light of ACT,
Phys. Lett. B 872 (2026) 140054,
arXiv:2507.06544 [astro-ph.CO],
doi:10.1016/j.physletb.2025.140054.

\bibitem{35'}
Z. Yi, X. Wang, Q. Gao, and Y. Gong,
Approximate reconstruction of inflationary potential with
ACT observations,
Phys. Lett. B 871, 140002 (2025),
doi:10.1016/j.physletb.2025.140002.

\bibitem{36'} S. N. Gashti et al., From minimal Higgs inflation with (R 2) term in palatini gravity to Swampland conjectures under ACT constraints, Eur. Phys. J. C 85.11 (2025) 1343.

\bibitem{37'}
M. A. S. Afshar et al.,
Swampland conjectures through ACT observations:
Observational signatures of radiative-corrected inflation,
Ann. Phys. 494, 170651 (2026),
doi:10.1016/j.aop.2026.170651.
    
\bibitem{Malekpour2024}
M. Malekpour, K. Nozari, F. Rajabi and N. Rashidi,
Non-minimal unimodular inflation,
Phys. Dark Univ. 43, 101405 (2024).

\bibitem{BK18} P. A. R. Ade et al. [BICEP/Keck Collaboration], BICEP/Keck XIII: Improved Constraints on Primordial Gravitational Waves using Planck, WMAP, and BICEP/Keck Observations through the 2018 Observing Season, Phys. Rev. Lett. 127, 151301 (2021), arXiv:2110.00483 [astro-ph.CO].
    
\bibitem{AndriotRoupec}
D. Andriot and C. Roupec,
Further refining the de Sitter swampland conjecture,
Fortschr. Phys. 67, 1800105 (2019),
arXiv:1811.08889 [hep-th].

\bibitem{LiuSWGC}
Y. Liu,
Higgs inflation and scalar weak gravity conjecture,
Eur. Phys. J. C 82, 1052 (2022),
doi:10.1140/epjc/s10052-022-10993-8.

\bibitem{Palti}
E. Palti,
The Weak Gravity Conjecture and Scalar Fields,
JHEP 08 (2017) 034,
arXiv:1705.04328 [hep-th].

\bibitem{GonzaloIbanez}
E. Gonzalo and L. E. Ibáñez,
A Strong Scalar Weak Gravity Conjecture and Some Implications,
JHEP 08 (2019) 118,
arXiv:1903.08878 [hep-th].

\bibitem{ACTSPTdiscussion}
R. Kallosh, A. Linde, and D. Roest,
Atacama Cosmology Telescope, South Pole Telescope, and Chaotic Inflation,
Phys. Rev. Lett. 135, 161001 (2025),
arXiv:2503.21030 [hep-th].
\end{thebibliography}
\end{document}